# Multiple Quantum Coherence and Entanglement Dynamics in Spin Clusters

G. B. Furman[1,2], V.M. Meerovich[1], and V.L. Sokolovsky[1]

[1]Department of Physics, Ben Gurion University, Beer Sheva 84105, Israel

[2]Ohalo College, Qazrin, 12900, Israel

25 May 2009

**Abstract**

With the purpose to reveal consistency between multiple quantum (MQ) coherences and entanglement, we investigate numerically the dynamics of these phenomena in one-dimensional linear chains and ring of nuclear spins 1/2 coupled by dipole–dipole interactions. As opposed to the calculation of the MQ coherence intensity based on the density matrix describing the spin system as a whole, we consider the "differentiated" intensity related only to the chosen spin pair based on the reduced density matrix. It is shown that the entanglement and the MQ coherence have similar dynamics only for nearest neighbors while we did not obtained any consistency for remote spins.

## I. Introduction

Entanglement is a central phenomenon of quantum mechanics [1-4] and plays a predominant role in quantum information processing applications such as quantum computing [5], quantum communication [6], and quantum metrology [7, 8]. At the last decade, entanglement appears as a major goal of many studies to be aimed for creation various quantum states with photons, trapped ions, cold atoms or spins. While being intensive study of the entanglement properties, including both qualitative and quantitative aspects [3, 4, 9], to decide whether a quantum state is entangled or not is still an unsolved problem in general.

For a spin system, which we consider in this paper, entanglement is regarded as a result of a quantum correlation between remote particles of the spin system [3, 4].

On the other hand, correlations between spins lead to the appearance of multiple-quantum (MQ) coherences [10]. Whereas MQ coherences is a collective effect of the correlation of all spins in the system [10, 11], widely used measures of entanglement, the von Neumann entropy and concurrence, describe entanglement between two spins only [3, 4]. Therefore, we will try to extract the intensity of the MQ coherence related to two selected spins in a spin cluster and to compare it with the concurrence of these spins.

For this purpose we will calculate the second and zero order coherences using the reduced density matrix as it is done for entanglement measure for a selected spin pair [12, 13]. Using the reduced density matrix we numerically simulate the MQ coherence and entanglement dynamics in one-dimensional linear chains and ring of nuclear spins 1/2 coupled by dipole–dipole interactions up to ten spins at low temperature.

## II. Entanglement measure and reduced density matrix

The most natural and widely used quantitative measures of entanglement is entanglement of formation [12], which is intended to quantify the resources needed to create an entangled state of two spins

$$E_F(\rho_{mn}) = Tr(\rho_{mn} \ln \rho_{mn}) \quad , \tag{1}$$

where $\rho_{mn}$ is the reduced density matrix, which describes dynamics of the $m$-th and $n$-th spins. For $m$-th and $n$-th spins, the reduced density matrix $\rho_{mn}$ defined by

$$\rho_{mn}(\tau) = Tr_{mn}(\rho(\tau)), \tag{2}$$

where $Tr_{mn}(...)$ denotes the trace over the degrees of freedom for all spins except the $m$-th and $n$-th ones, $\tilde{\rho}_{mn}(\tau)$ is the complex conjugation of $\rho_{mn}$. The analytic expression for $E_F$ is given by

$$E_F(x) = -x\log_2 x - (1-x)\log_2(1-x) , \tag{3}$$

where $x = \tfrac{1}{2}\left(1+\sqrt{1-C^2}\right)$ and $C$ is the concurrence between two spins [12]. For maximally entangled states, the concurrence is $C=1$ while for separable states $C=0$. The concurrence between the two spins $m$ and $n$ is expressed by the formula

$$C_{mn} = \max\left\{0,\ 2\lambda_{mn} - \sum_{k=1}^{4}\lambda_{mn}^{(k)}\right\} , \tag{4}$$

where $\lambda_{mn} = \max\{\lambda_{mn}^{(k)}\}$ and $\lambda_{mn}^{(k)}(k=1,2,3,4)$ are the square roots of the eigenvalues of the product

$$R_{mn}(\tau) = \rho_{mn}(\tau)(\sigma_y \otimes \sigma_y)\tilde{\rho}_{mn}(\tau)(\sigma_y \otimes \sigma_y). \tag{5}$$

Here $\sigma_y = \begin{pmatrix} 0 & -i \\ i & 0 \end{pmatrix}$ is the Pauli matrix.

## III. MQ NMR dynamics and two-spin coherences at low temperature

Let us consider an MQ NMR experiment on a system of nuclear spins coupled by the dipole-dipole interaction (DDI) in a strong external magnetic field $\vec{H}_0$ at low temperature. At initial time $t=0$ the spin system is assumed to be in thermal equilibrium with the lattice and the equilibrium spin density operator $\rho_{eq}$ has the

following form

$$\rho_{eq} = \frac{\exp(-\beta H)}{Tr[\exp(-\beta H)]} \quad (6)$$

where $H$ is the Hamiltonian of the system, $\beta = \frac{\hbar}{kT}$, $T$ is the Zeeman temperature.

The basic scheme of MQ NMR experiments consists of four distinct periods of time: preparation, evolution, mixing and detection [10]. There are many radiofrequency (RF) pulse sequences exciting MQ coherences during the preparation period. For a dipolar-coupled spin system, the multiple pulse sequence with an eight-pulse cycle is known to be very efficient [10]. It creates the double-quantum effective Hamiltonian

$$H_{MQ} = H^{(2)} + H^{(-2)}, \quad (7)$$

where $H^{(\pm 2)} = -\frac{1}{4}\sum_{j<k} D_{jk} I_j^{\pm} I_k^{\pm}$, $D_{jk} = \frac{\gamma^2 \hbar}{r_{jk}^3}(1 - 3\cos^2\theta_{jk})$ is the coupling constant between spins $j$ and $k$, $\gamma$ is the gyromagnetic ratio, $r_{jk}$ is the distance between spins $j$ and $k$, $\theta_{jk}$ is the angle between the internuclear vector $\vec{r}_{jk}$ and the external magnetic field $\vec{H}_0$ which is directed along the z-axis, $I_j^+$ and $I_j^-$ are the raising and lowering operators of spin $j$.

The density matrix of the spin system, $\rho(\tau)$, at the end of the preparation period is

$$\rho(\tau) = U(\tau)\rho_{eq}U^+(\tau), \quad (8)$$

where $U(\tau) = \exp(-i\tau H_{MQ})$. Then the evolution period without any pulses follows. The transfer of the information about MQ coherences to the observable magnetization occurs during the mixing period. The resulting signal $S(\tau,t)$ stored as population information reads [10, 14]

$$S(\tau,t) = \sum_k e^{-ik\Delta\omega t} J^k(\tau) \quad (9)$$

where $\Delta\omega$ is the RF offset, chosen to be larger than the local dipolar field frequency, $\omega_d$ [10]. $J_k(\tau)$ is the spectral intensities of order $k$ [14]

$$J_k(\tau) = Tr[\rho(\tau)\rho^{zk}(\tau)] \quad (10)$$

Here $\rho^{zk}(\tau)$ is determined in the following way: first we calculate $\rho_z(\tau) = U(\tau)I_z U^+(\tau)$ ($I_z$ is the z-component of the spin angular momentum operator) and then the terms of $\rho_z(\tau)$ are grouped according to their MQ order $k$: $\rho_z(\tau) = \sum_k \rho^{zk}(\tau)$ [10, 14].

The intensities given by (10) are integrated characteristics describing MQ coherences of the order $k$ in a spin system. At the same time, the von Neumann entropy and concurrence describe entanglement between two selected spins $m$ and $n$. It would be more adequate to compare correlations between spins, which have the direct attitude to entanglement and correlations between the same spins which lead to occurrence of MQ coherences. One could expect that the spectral intensity of the MQ coherence of the second order given by the reduced density matrix characterizes all the possible coherences in which the spins $m$ and $n$ are involved. Therefore, we will use the same reduced density matrix $\rho_{mn}$ to calculate both the MQ coherence intensities and concurrence between two spins in a multi-spin system.

The spectral intensities $J_k^{mn}(\tau)$ of the coherences of two spins $m$ and $n$ can be determined using the reduced density matrix (5) as:

$$J_k^{mn}(\tau) = Tr\left[\rho_{mn}(\tau)\rho_{mn}^{zk}(\tau)\right], \qquad (11)$$

where $\rho_{mn}^{zk}(\tau) = Tr_{mn}\left(\rho^{zk}(\tau)\right)$. Obviously, that the configuration of MQ coherences (11) consists of the coherences of the zeroth and second orders only.

## VI. Two-spin MQ dynamics and evolution of entanglement

We will study a spin system which is initially in thermodynamic equilibrium state at low temperature and is described by density matrix (6). We restrict ourselves to numerical simulations of MQ and entanglement dynamics for one-dimensional and circular (ring) spin systems. Ring of six dipolar-coupled proton spins of a benzene molecule [15], hydroxyl proton chains in calcium hydroxyapatite $Ca_5(OH)(PO_4)_3$ [16] and fluorine chains in calcium fluorapatite $Ca_5F(PO_4)_3$ [16] are examples of suitable objects to study MQ dynamics by NMR technique. In our numerical simulations, the dipolar coupling constant of the nearest neighbors is chosen to be $D_{j,j+1} = 1\,s^{-1}$. We assume also that the angles $\theta_{jk}$ are the same for all pairs of spins and the distances between nearest neighbors $r_{jk}$ are equal. Then the coupling constants of spins $j$ and $k$ are $D_{j,j+1}\left[\frac{\sin(\pi/N)}{\sin(\pi(j-k)/N)}\right]^3$ for the ring and $D_{j,j+1}/|j-k|^3$ for the chain, respectively. The numerical simulations of the MQ and entanglement dynamics are performed using the software based on the MATLAB package, allowing us to investigate spin systems up to ten spins. Along with evolution of the MQ coherences, $J_k(\tau)$, given by (10), we examine the time dependence of the two-spin coherences, $J_k^{mn}(\tau)$, given by (11), and concurrence, $C_{mn}(\tau)$ given by (4), between the spins $m$ and $n$ when the spin system evolves under the Hamiltonian $H$. In our calculations, the parameter $\beta\|H\|$ (here $\|...\|$ denotes a norm of the operator $H$) which determines the temperature dependence of intensities of MQ coherences is taken 10. For protons in the external magnetic field $H_0 = 5\,\text{T}$, this value corresponds to the temperature of 1 mK. Fig. 1 shows time dependence of the "integrated" $J_2$ and "differentiated" $J_2^{mn}$ intensities of the MQ coherences, and concurrences $C_{mn}$ between various spin pairs in the four spin chain. The initial period of evolution is characterized by the creation of entanglement only between the nearest neighbors in the chain (Fig. 1a). Then, entanglement develops between distant spins (Fig. 1b and c). The longer is the distance between spins, the more time the appearance of their entanglement takes. One can see from Fig. 1a that the concurrence $C_{1,2}$ and the spectral intensity $J_2^{1,2}$ have qualitatively close dynamics. They reach their maximal and minimal values at the same moments of time. Similar behavior was found for all nearest neighbors. This dynamics consistency with the large stretch can be attributed to concurrence and spectral intensity of the next-next neighbors, but in any way it is impossible to attribute to next neighbors. It is worth noticing that the results of the numerical calculations show no consistency between concurrence and integrated spectral intensity $J_2$ (Fig. 1).

The results of the numerical calculation of the time dependence of the concurrences and spectral intensities in a circle of six dipolar-coupled spins and in a chain of ten spins are presented in Figs 2 and 3, respectively. One can see from Figs. 2a and 3a that there is consistency between spectral intensities of the coherences of two spins $J_2^{1,2}$ and concurrencies $C_{1,2}$ for the nearest neighbors while these characteristics for remote spins do not show any consistency (Figs 2b, 2c and 3b). The same conclusion can be done for any spin pairs.

Note one interesting feature in behavior of the concurrence between next neighbors ($C_{1,3}$) in a six spin circle: the entangled quantum states arise practically at the initial stage of evolution (Fig. 2b) and grow more rapidly than the intensities $J_2$ and $J_2^{1,3}$. For ten spin chains, entanglement between the first spin and spins located at the middle of the chain disappears, in spite of the direct interaction between these spins. Similar results have been obtained for the 8- and 9- spin chains [17]. Surprisingly, that, at the same time, the concurrences $C_{1,10}$ between the ends of the chain, i.e. between the most remote spins, are non-zero (Fig. 3b).

## Conclusions

In order to adequately compare entanglement between different spins and MQ coherences in a spin system, we have proposed to use the differentiated characteristic of the coherences between chosen spin pairs and calculate the differentiated intensities using the reduced density matrix. Our numerical calculations show that the consistency between spin coherence and entanglement appears only for the nearest neighbors while for remote spins no consistency was observed. Since the chosen spin pair is always involved into formation of coherence intensity of the highest order of MQ coherence (all the spins of a system change their state simultaneously), it is interesting to analyze consistency between this coherence and the concurrence. Such consistency is observed only for of the most remote spins in the ring of six spins for times greater than $\tau = \dfrac{5}{D_{1,2}}$ (sixth order coherence intensity $J_6$ is shown by the blue dash-doted line in Fig. 2c). However, for a in one-dimensional linear ten-spin chain, the consistency of the highest order of MQ coherence possessing the intensity $J_{10}$ with concurrence $C_{1,10}$ is displayed only for their first extremums (Fig. 3c). Thus, even the differentiated approach for calculation of the

MQ coherence intensity corresponding to the chosen remote spin pair is not revealed any general consistency between dynamics of the coherence and entanglement.

References

1. M. A. Nielsen and I. L. Chuang, Quantum Computation and Quantum Information (Cambridge University Press, Cambridge, 2000).
2. G. Benenti, G. Casati, and G. Strini, Principles of Quantum Computation and Information, Volume I and II, (World Scientific, 2007).
3. L. Amico, R. Fazio, A. Osterloh, and V. Vedral, Rev. Mod. Phys. **80**, 517 (2008).
4. R. Horodecki, P. Horodecki, M. Horodecki, K.Horodecki, http://arxiv.org/abs/quant-ph/0702225v1.
5. C. H. Bennett and D. P. DiVincenzo, Nature **404**, 247--255 (2000).
6. C. H. Bennett, G. Brassard, C. Crepeau, R. Jozsa, A. Peres, W.K. Wootters, Phys. Rev. Lett. **70**, 1895 (1993).
7. C. F. Roos, K. Kim, M. Riebe, R. Blatt, Nature, **443**, 316 (2006).
8. P. Cappellaro, J. Emerson, N. Boulant, C. Ramanathan, S. Lloyd, and D. G. Cory, Phys. Rev. Lett., **94**, 020502 (2005).
9. T. Konrad, F. de Melo, M. Tiersch, C. Kasztelan, A. Aragão, A. Buchleitner, Nature Physics **4**, 99 (2008).
10. J. Baum, M. Munovitz, A. N. Garroway, A. Pines, J. Chem. Phys. **83**, 2015 (1985).
11. J.Tang and A.Pines, J.Chem.Phys., **73**, 2512 (1980).
12. W. K.Wootters, Phys. Rev. Lett **80**, 2245 (1998)
13. C. H. Bennett, D. P. DiVincenzo, J. A. Smolin, and W. K. Wootters, Phys. Rev. A **54**, 3824, (1996)
14. E. B. Fel'dman and I. I. Maximov, J. Magn. Reson. **157**, 106 (2002).
15. J.-S. Lee and A. K. Khitrin, Phys. Rev. A, **70**, 022330 (2004)
16. G. Cho, J. P. Yesinowski, J. Phys. Chem. **100**, 15716 (1996)
17. G. B. Furman, V. M. Meerovich, and V. L. Sokolovsky, Phys. Rev. A, **78**, 042301 (2008).

Figures

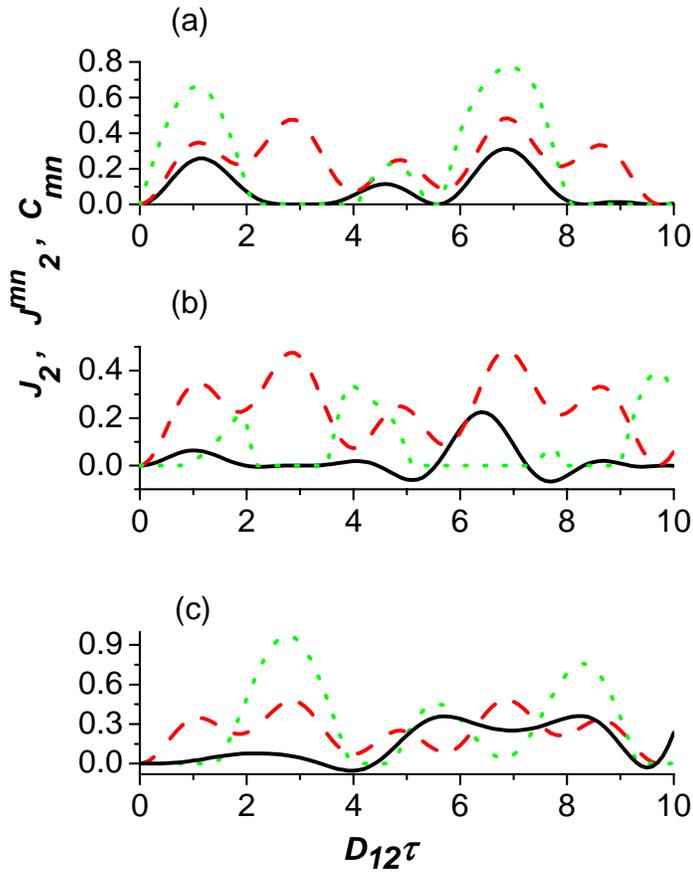

Fig. 1 (Color online) Time dependence of the differentiated $J_2^{mn}$ (black solid line) and integrated $J_2$ (red dashed line) intensities of the MQ coherences, and concurrences $C_{mn}$ (green dotted line) in a four-spin chain. (a) m=1, n=2; (b) m=1, n=3, $J_2^{1,3} \times 20$; (c) m=1, n=4, $J_2^{1,4} \times 10$.

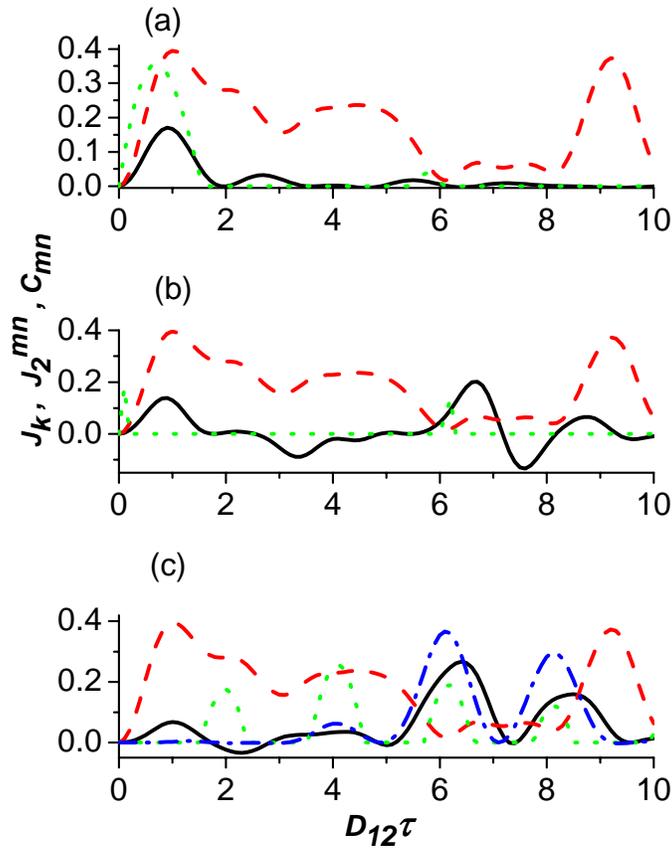

Fig. 2 (Color online) Time dependence of the differentiated $J_2^{mn}$ (black solid line) and integrated $J_2$ (red dashed line) intensities and concurrences $C_{mn}$ (green dotted line) in the six spin circle. (a) m=1, n=2; (b) m=1, n=3, $J_2^{1,3} \times 20$; (c) m=1, n=4, $J_2^{1,4} \times 10$. Blue dash-dotted line shows intensity $J_6$ of the MQ coherence of the sixth order.

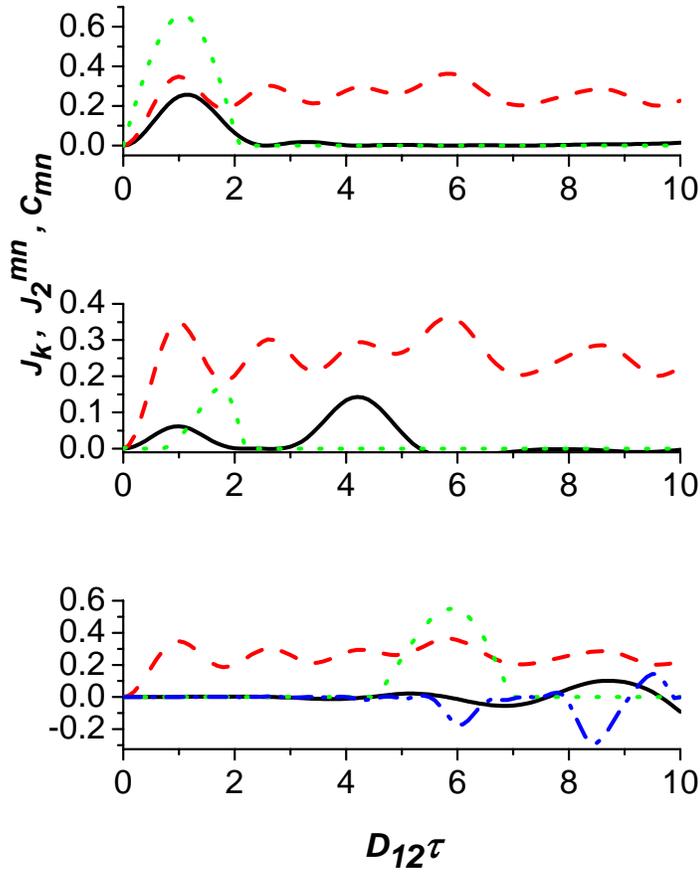

Fig. 3 (Color online) Time dependence of the differentiated $J_2^{mn}$ (black solid line) and integrated $J_2$ (red dash line) intensities of the MQ coherences and concurrences $C_{mn}$ (green dotted line) in the ten spin chain. (a) m=1, n=2; (b) m=1, n=3, $J_2^{1,2} \times 20$; (c) m=1, n=10, $J_2^{1,10} \times 10^3$. Blue dash-dotted line shows intensity $J_{10} \times 10^3$ of the MQ coherence of the tenth order.